\documentclass[twocolumn,aps,pre,showpacs,amsmath,amssymb]{revtex4}

\usepackage[dvips]{graphicx}
\usepackage{dcolumn}
\usepackage{bm}
\newcommand{\Rv}{{\mathbf R}}
\newcommand{\rv}{{\mathbf r}}

\newcommand{\hv}{{\mathbf h}}
\newcommand{\xv}{{\mathbf x}}

\newcommand{\qv}{{\mathbf q}}

\newcommand{\Tr}{{\rm Tr}}
\newcommand{\px}{{\partial_x}}
\newcommand{\py}{{\partial_y}}

\def\mm#1{ \underline{\underline{{#1}}}}
\def\gm{{\mm{g}}}
\def\Qm{{\mm{Q}}}

\def\Im{{\mm{I}}}
\def\um{{\mm{u}}}

\begin{document}
\preprint{APS/123-QED}
\title{Phases and Transitions in Phantom Nematic Elastomer Membranes}
\author{Xiangjun Xing}
\affiliation{Department of Physics, University of Illinois,
  Urbana-Champaign, IL 61801}
\author{Leo Radzihovsky}
\affiliation{Department of Physics, University of Colorado, Boulder,
  CO 80309}%

\date{\today}

\begin{abstract}
  Motivated by recently discovered unusual properties of bulk nematic
  elastomers, we study a phase diagram of liquid-crystalline
  polymerized phantom membranes,
  focusing on in-plane nematic order. We predict that such membranes
  should generically exhibit five phases, distinguished by their
  conformational and in-plane orientational properties, namely
  isotropic-crumpled, nematic-crumpled, isotropic-flat, nematic-flat
  and nematic-tubule phases. In the nematic-tubule phase, the membrane
  is extended along the direction of {\em spontaneous} nematic order
  and is crumpled in the other. The associated spontaneous symmetries
  breaking guarantees that the nematic-tubule is characterized by a
  conformational-orientational soft (Goldstone) mode and the
  concomitant vanishing of the in-plane shear modulus. We show that
  long-range orientational order of the nematic-tubule is maintained
  even in the presence of harmonic thermal fluctuations. However, it
  is likely that tubule's elastic properties are qualitatively
  modified by these fluctuations, that can be studied using a
  nonlinear elastic theory for the nematic tubule phase that we derive
  at the end of this paper.
\end{abstract}

\pacs{PACS:61.41.+e, 64.60.Fr, 64.60.Ak, 46.70.Hg}
\maketitle
\section{Introduction}
\label{sec:intro}
Fluctuating polymerized\cite{polymerized_comment}
membranes\cite{membrane_JWS} have attracted considerable interests in
the past two decades. Probably the most striking property that
distinguishes them from their one-dimensional polymer analogue and
most other two-dimensional systems, is that polymerized membranes
admit a low-temperature flat
phase\cite{NelsonPeliti,membrane_AL,membrane_LR}
characterized by a long-range order in membrane's normal. This phase
is separated from a high-temperature crumpled phase, with
randomly-oriented normals, by a thermodynamically-sharp
transition\cite{membrane_KN,membrane_PKN,membrane_GD}.  At low
temperatures, strong interplay between thermal fluctuations and
elastic nonlinearities infinitely enhances the effective bending
rigidity, which in turn stabilizes the flat phase against these very
fluctuations\cite{JWS_radzihovsky,NelsonPeliti,membrane_AL,membrane_GD}.
The resulting flat phase exhibits rather unusual length-scale
dependent elasticity, nonlinear response to external stress and a
universal negative Poisson ratio.\cite{membrane_AL,membrane_LR}

More recently, stimulated by a number of possible anisotropic
realizations, the role of elastic membrane anisotropy was considered
by Radzihovsky and Toner\cite{RTtubules}.
It was discovered that while permanent in-plane anisotropy is unimportant
at long scales within the crumpled and flat phases, it qualitatively
modifies the global phase diagram of anisotropic polymerized
membranes, leading to an intermediate (in its properties and location
in the phase diagram) ``tubule'' phase.  In the tubule phase the
membrane is extended in one direction and crumpled in the other, with
its crumpled boundary exhibiting statistics of a self-avoiding
polymer. The usual crumpled-to-flat transition of an isotropic
membrane is therefore generically split into two consecutive ones:
crumpled-to-tubule and tubule-flat transitions.  Fluctuations,
nonlinear elasticity, and phase transitions into and out of the tubule
phase have been studied extensively\cite{RTtubules}, with predictions
dramatically confirmed by large-scale Monte Carlo
simulations\cite{membrane_BFT}.

Motivated by unusual properties of nematic elastomers that have been
recently discovered\cite{GolLub89,review_exp,review_theory,
stress_strainExps,elast_us,Xing-Radz,StenLub}, i.e. a spontaneous
nematic order in an amorphous solid phase and resulting soft
elasticity characterized by one vanishing shear modulus,
we have previously studied fluctuations and elasticity of a flat tethered
membrane that exhibits {\em spontaneous} in-plane anisotropy, i.e. a nematic
order\cite{membrane_nematicflat}. Such spontaneous order guarantees
the existence of a zero-energy in-plane deformation mode corresponding to
a simultaneous reorientation of nematic and uniaxial distortion axes,
and hence vanishing of an in-plane shear modulus. As in other ``soft''
fluctuating systems\cite{comment_soft},
we found that elastic nonlinearities become qualitatively important
in determining membrane's long-scale elasticity and conformations.
However, due to the vanishing of an in-plane shear modulus,
the elasticity of flat nematic elastomer membranes differ
qualitatively from that of isotropic and that of explicitly
anisotropic membranes: a renormalization analysis of the undulation
nonlinearity shows that nematic-flat phase is characterized by a
conventional, length scale independent bending-rigidity
modulus.\cite{membrane_nematicflat,comment_nematicflat}

While bulk nematic elastomers and gels have received considerable
experimental and theoretical attention, properties of equally
interesting two-dimensional system, i.e.  nematic elastomer membranes,
have, to the best of our knowledge have not been explored in any
detail.  A few (rather idealized) experimental realizations of such
{\em spontaneously} anisotropic nematic membranes can be envisioned.
For example, one may prepare a nematic elastomer sheet in its
isotropic phase and lower the temperature of the elastomer membrane
into its nematic state.  The chemical structure of liquid crystalline
mesogenic units in such nematic elastomer should be chosen such that
they prefer alignment parallel to the membrane surface.
Liquid-crystalline polymers adsorbed onto a polymerized membrane can
also develop a spontaneous nematic order at sufficiently high
density\cite{Podgornik}, which in turn will act as a spontaneous
elastic anisotropy to the membrane elasticity.  The elastic properties
of lipid bilayers with spontaneous tilt order should be almost
identical to nematic elastomer membranes that we study in this
paper\cite{comment_defects}. Finally, other types of liquid
crystalline orders may also be realized in polymerized membranes. For
example, two dimensional smectic order may spontaneously develops when
temperature is further lowered in a nematically-ordered elastomer
membrane. This can also be taken into account using a straightforward
generalization of our model.

\begin{figure}
\begin{center}
\includegraphics[width=6.5cm]{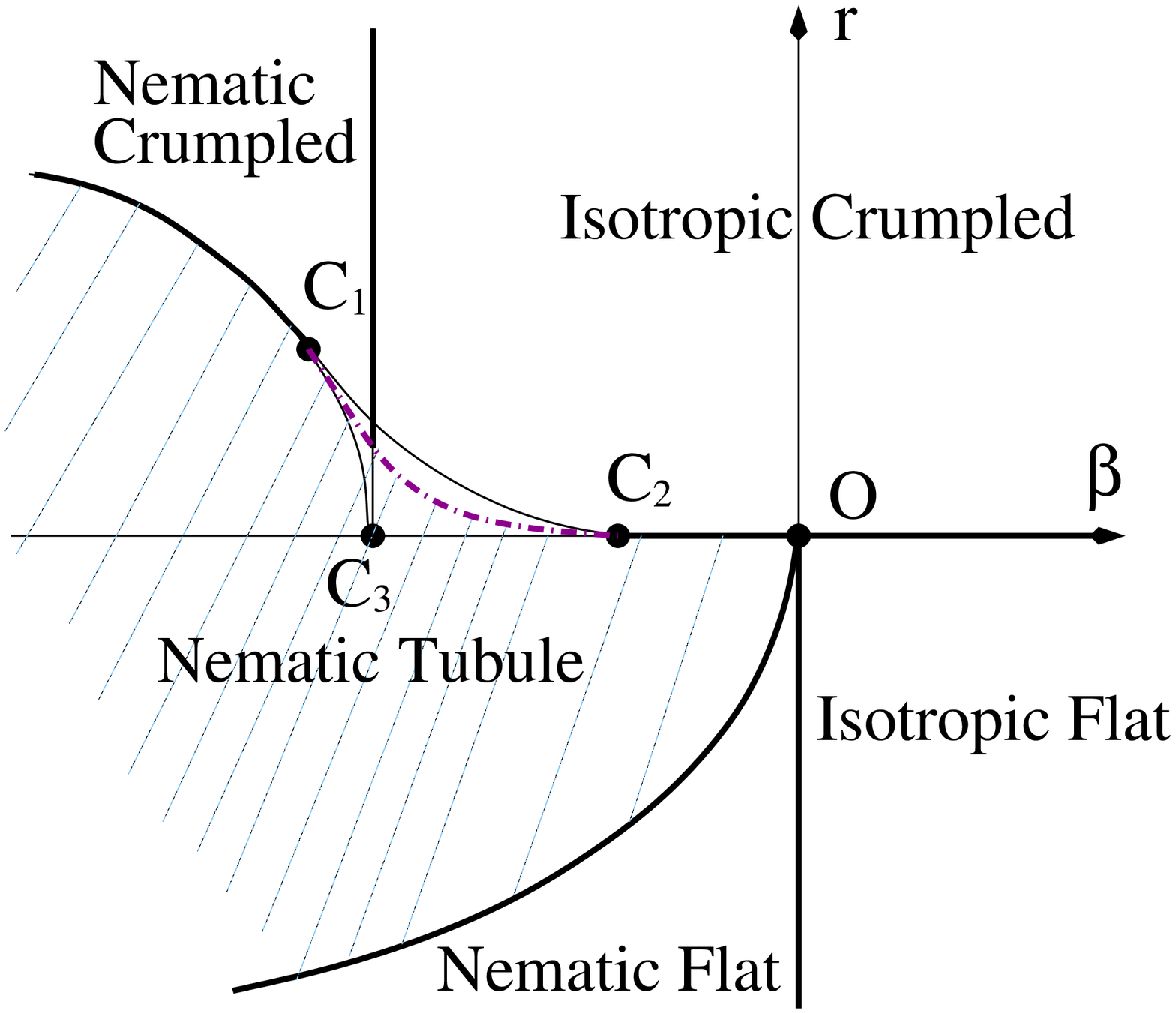}
\caption{(Schematic mean-field phase diagram for a polymerized
  membrane with spontaneous in-plane nematic order.  Parameters $r$
  and $\beta$ drive crumpling transition and isotropic-nematic
  transition respectively and are defined in Sec.\ref{sec:model} and
  Sec.~\ref{sec:phasediagram}.  Nematic-tubule phase is the stable
  phase in the shaded region.  In the triangle region between three
  critical points $C_1$, $C_2$ and $C_3$, two phases are metastable.
  The dash line connecting $C_1$ and $C_2$ is a (schematic)
  first-order transition line between isotropic/nematic-crumpled
  phases and the nematic-tubule phase. }
\label{phasediagram}
\end{center}
\end{figure}

In this paper we continue to explore the physics of
phantom\cite{comment_phantom} liquid-crystalline tethered membranes,
focusing on the overall phase diagram that results from the interplay
of membrane's in-plane nematic and conformational orders.  As
discussed in detail in Sec.~\ref{sec:phasediagram}, we find that in
the isotropic sector of the phase diagram, characterized by a
vanishing in-plane nematic order, the system is identical to the usual
isotropic tethered membranes, and therefore exhibits well studied
isotropic-crumpled (IC) and isotropic-flat (IF) phases.
Development of nematic order generically leads to three additional
phases: the nematic-crumpled (NC), nematic-tubule (NT) and
nematic-flat (NF) phases. Different phases can be distinguished by
different expectation values of the nematic order parameter and the
metric tensor.  These results are summarized by a mean-field global
phase-diagram in Fig.~\ref{phasediagram}.

\begin{figure}
\begin{center}
\includegraphics[width=6.5cm]{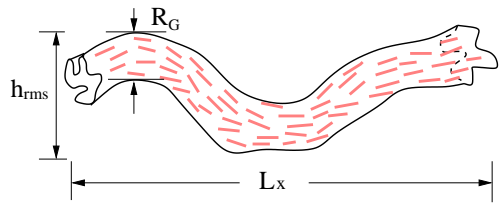}
\caption{Conformational properties of the nematic tubule phase.
  Average thickness $R_G$ and root-mean fluctuation $h_{\mbox{rms}}$
  are defined in Eqs.~(\ref{Rg_h}).  Small rods, aligned on the
  membrane's surface schematically indicate existence of the in-plane
  nematic order.}
\label{nematic_tubule}
\end{center}
\end{figure}

We also study harmonic fluctuations within the most interesting
nematic-tubule phase.  As we discuss in Sec.~\ref{sec:tubule} and show
schematically in Fig.~\ref{nematic_tubule}, the geometry and
fluctuations of a nematic tubule, with an intrinsic size $L\times L$,
are characterized by its average thickness $R_G(L)$ and root-mean
fluctuations $h_{rms}(L)$ about its energetically preferred extended
state, that scale as:
\begin{equation}
R_G(L) \propto L^{\nu}, \hspace{5mm}
h_{\mbox{rms}}(L) \propto L^{\theta},
\end{equation}
where $\nu$ and $\theta$ are universal exponents.\cite{RTtubules} For
idealized ``phantom'' membranes $\nu = 1/4$ and $\theta = 1$, but are
expected to be significantly modified by the self-avoidance and
elastic nonlinearities.\cite{RTtubules} These conformational
properties (or the absence thereof) can also be used as indicators of
phase transitions into/out-of the nematic tubule phase, accompanied by
usual thermodynamic singularities, e.g., in the heat-capacity. For a
more detailed discussion of the scaling properties of polymerized
tubules and transitions into/out-of the tubule phase, we refer the
reader to Ref.\onlinecite{RTtubules}.

We also derive a fully rotationally invariant nonlinear elasticity
characterizing nematic tubules.  We find that because of the
spontaneous nature of the in-plane nematic order, the nonlinear
elastic free energy of the nematic tubule differs qualitatively from
the tubule phase of a permanently anisotropic
membrane\cite{RTtubules}.  We leave the full analysis of tubule's
elastic nonlinearities and thermal fluctuations to a future research.
Here, we will have little to say on the very interesting and important
question of the effects of heterogeneity, associated with the random
cross-linking of elastomers.  Based on a recent body of work on
conventional liquid crystals confined in random
environments\cite{RTaerogels,RT_MSC} and randomly polymerized
membranes\cite{membrane_random}, we expect that at sufficiently long
scales, even weak heterogeneities will lead to qualitatively important
modifications of some of the predictions made in this paper.
Understanding these will be essential for a direct comparison with
experiments\cite{Safinya_Unpublished}.  We plan to explore these in
the future.

The body of this paper is organized as follows.  In Section
\ref{sec:model} we introduce a Landau model for a liquid-crystalline
tethered membrane, whose mean field phase diagram we map out in
Section \ref{sec:phasediagram}.  In Section \ref{sec:tubule}, we focus
on the most interesting phase, the nematic-tubule and study its
thermal fluctuations at the harmonic
level. 
In Section \ref{sec:tubule_nl}, we formally integrate out the nematic
order parameter and derive an effective fully {\em nonlinear} and
rotationally-invariant elastic free energy for the nematic tubule phase.

\section{Model of Nematic Elastomer Membrane}
\label{sec:model}
Nematic elastomer membranes are direct generalization of permanently
anisotropic tethered membranes studied in
Ref.~[\onlinecite{RTtubules}], with the in-plane elastic
anisotropy chosen by a {\em spontaneously}, local nematic order.  The
latter plays the role of a new degree of freedom that fluctuates at a
finite temperature. As for an ordinary tethered membrane, its
geometric conformation (state) is described by a three-dimensional
vector-field $\Rv(\xv)$, that specifies the position (embedding) of
the reference-space mass point $\xv$ in the three dimensional target
space. Throughout this paper, we will use a lower-case bold-face
characters, such as $\xv= x_a \hat{e}_a = x \hat{x} + y\hat{y}$, to
denote two-dimensional vectors in the reference space and upper-case
bold-face character such as $\Rv=X \hat{X} + Y \hat{Y} + Z \hat{Z}$ to
denote three-dimensional vectors in the target space, respectively
spanned by a set of mutually orthogonal unit vectors, $\{ \hat{e}_1
=\hat{x}, \hat{e}_2 = \hat{y} \}$ and $\{ \hat{X}, \hat{Y},
\hat{Z}\}$.

The elastic free energy density of the membrane can be conveniently
described as a function of a reference-space metric tensor $\mm{g}$
\begin{equation}
g_{ab} = 
\frac{\partial \Rv}{\partial x_a}\cdot \frac{\partial \Rv}{\partial x_b},
\label{def_g}
\end{equation}
inherited from the Euclidean metric $\delta_{ij}$ of the embedding
space.  By construction, $g_{ab}$ is explicitly a target-space scalar
(i.e., rotationally invariant in the embedding space), rank-2
reference-space tensor, and is non-negative.  The thermal expectation
value of the metric tensor, $\langle \gm\rangle$ can be used to
distinguish different phases of a tethered membrane. More specifically,
two eigenvalues of $\langle\mm{g}\rangle$ describe whether and how the
membrane is extended along two directions in its reference space.  
For example, $\langle\mm{g}\rangle$ vanishes in the crumpled phase, 
and is proportional to the identity matrix in the flat phase, with
the common eigenvalue describing the relative size of the flat 
membrane in the three dimensional target space. 
In the tubule phase of an intrinsically anisotropic membrane,
$\langle \gm\rangle$ has only one positive eigenvalue, which
corresponds to the fact that the membrane is extended along one
direction and crumpled along others. Therefore $\langle\mm{g}\rangle$
has all the properties of a physical order parameter.

Nematic elastomer membranes are in addition characterized by a local
in-plane nematic order parameter field $\mm{Q}(\xv)$. Characterizing
membrane's spontaneous elastic anisotropy, $\mm{Q}$ describes an intrinsic
property of the membrane and is a second-rank symmetric traceless tensor 
in the two dimensional reference space, with components $Q_{ab}$, 
$a,b=x,y$ and $\sum_a Q_{aa} = 0$, and is a target-space 
scalar\cite{comment_Q}.
Landau free energy density for our model of
a nematic elastomer membrane is a rotationally-invariant (both in the
reference and target spaces) function of $\mm{g}$, $\mm{Q}$:
\begin{equation}
\tilde{f}[\gm,\Qm] = f_e[\gm] + f_n[\Qm] + f_{en}[\gm, \Qm],
\label{free_energy_0}
\end{equation}
where $f_e$ is the elastic free energy density for an
isotropic polymerized membrane\cite{membrane_PKN},
\begin{equation}
f_e[\gm] = \frac{\kappa}{2}(\nabla^2 \Rv)^2 + r' \Tr \gm +
\frac{\lambda}{2} (\Tr \gm)^2 + \mu\Tr \gm^2\label{elastic}.
\end{equation}
$f_n$ is the Landau-deGennes free energy density for a nematic-isotropic
transition in two dimensions
\begin{equation}
f_n = \frac{K}{2}(\nabla \Qm)^2 + \frac{\beta'}{2}\Tr \Qm^2 +
\frac{v}{2}(\Tr \Qm^2)^2.\label{nematic}
\end{equation}
It reflects the special feature of nematic-isotropic transition in two
dimensions, namely, the absence of the usual cubic invariant $\Tr
\Qm^3$ that vanishes identically. This allows nematic-isotropic
transitions in two dimensions to be continuous\cite{LC_deGennes}.

The last part of $\tilde{f}$, Eq.~\ref{free_energy_0} is the coupling
between nematic order and the elastic degrees of freedom, which, to
lowest order, is given by:
\begin{equation}
f_{en} = - \alpha \Tr\,\Qm \gm = - \alpha \Tr\,\Qm \tilde{\gm},
\label{coupling}
\end{equation}
where $\tilde{\gm} = \gm - (Tr \gm)\, \mm{I}/2$ is the traceless part
of the metric tensor. In the presence of such tensorial nemato-elastic
coupling, nematic order $\Qm$ acts as a uniaxial shear stress and
therefore renders the in-plane elasticity anisotropic inside a nematic
phase. Concomitantly, an elastic distortion characterized by an
anisotropic metric tensor acts as a quadrupolar field inducing nematic
order. As usual in Landau description, although other higher order
couplings exist, such as for example $\gamma \Tr \gm \Tr \Qm^2$, they
do not lead to any new qualitative effects, and, furthermore, for weak
nematic order, are quantitatively subdominant. Their primary effect is
to slightly change the locus of phase boundaries without changing
their topology. We will therefore ignore them in this paper.

We can also formally and perturbatively integrate out the nematic
order parameter $\Qm$ in Eq.~(\ref{free_energy_0}) and obtain an
effective elastic free energy density formulated purely in terms of
the metric tensor. Up to fourth order in $\gm$, it is given by
\begin{eqnarray}
f_{\rm eff}[\gm] &=& \frac{1}{2} \kappa (\nabla^{2} \rv)^2 + 
\frac{1}{2} t \Tr \mm{g}
  + u (\Tr \tilde{\mm{g}}^2) + v (\Tr \mm{g})^2 \nonumber\\
&+& B (\Tr \tilde{\mm{g}}^2)^2 + C \Tr g \Tr \tilde{\mm{g}}^2,
\end{eqnarray}
and agrees with the purely elastic model used in our previous study of
the nematic-flat phase\cite{membrane_nematicflat}. Although such
effective model can also be used to study the phase diagram and other
phases, it cannot distinguish the nematic-crumpled and the
isotropic-crumpled phases, since both are characterized by a vanishing
metric tensor.  Therefore, depending on the equilibrium value of the
metric tensor, controlled by model parameters, $f_{\rm eff}$ predicts
a crumpled phase, a spontaneous tubule phase (nematic-tubule phase),
an isotropic-flat phase, and a spontaneous anisotropic flat phase
(nematic-flat phase).  Here, the usual crumpled-to-flat transition is
determined by changing the sign of $t$, while the crumpled-to-tubule
transition is driven by a sign of $u\propto\mu$. This purely elastic
approach has the advantage of simplicity, stemming from absence of the
nematic order parameter. In Sec.\ref{sec:tubule}, we will use this
effective elastic free energy to study the nonlinear elasticity in the
nematic-tubule phase.

\section{Mean-field Phase Diagram}
\label{sec:phasediagram}
An isotropic polymerized membrane with elastic free energy density,
Eq.~(\ref{elastic}) undergoes a continuous\cite{comment_crumpling}
crumpling transition as $r'$ is tuned from positive to
negative\cite{membrane_PKN}. On the other hand, $f_n$, Eq.~(\ref{nematic})
exhibits a second-order isotropic-nematic transition as $\beta'$
changes sign\cite{LC_deGennes}. When both $r'$ and $\beta'$ are small, a
competition between these two types of order becomes important.

To work out the mean-field phase diagram, we minimize the free energy
density Eq.~(\ref{free_energy_0}) over the metric tensor $\gm$ and the
nematic order parameter $\Qm$. To this end, it is convenient to
rescale both order parameters $\gm$ and $\Qm$ as well as the total
free energy Eq.~\ref{free_energy_0} as following,
\begin{subequations}
\begin{eqnarray}
\Qm &\rightarrow& \frac{\alpha}{4 \sqrt{\mu\,v}}\,\Qm,\\
\gm &\rightarrow& \frac{\alpha^2}{16\sqrt{\mu^3\,v}}\,\gm,\\
\tilde{f} &\rightarrow& \frac{\alpha^4}{256 \mu^2 \,v}\,f.
\end{eqnarray}
\end{subequations}
Furthermore, it is clear that the free energy 
is minimized by uniform nematic order and metric tensor. Therefore
we can ignore the two higher derivative terms $\kappa$, $K$ appearing
in Eqs.~(\ref{elastic}) and (\ref{nematic}), respectively. The
resulting rescaled total free energy density has the following simple
form:
\begin{eqnarray}
f[\gm,\Qm] &=& 
\frac{\epsilon}{2} (\Tr (\gm + r \Im -\Qm))^2 +  \Tr (\gm + r \Im -\Qm)^2
\nonumber\\
&+& \frac{1}{2} (\Tr \Qm^2 + \beta )^2. \label{model}
\end{eqnarray}
where
\begin{subequations}
\begin{eqnarray}
r &=&\frac{8\,r'\sqrt{\mu^3\,v}}{\alpha^2(\lambda+\mu)},\\
\beta &=& \frac{8\mu\beta'}{\alpha^2}-1,\\
\epsilon &=& \frac{\lambda}{\mu}, \label{def_epsilon}
\end{eqnarray}
\end{subequations}
and $\Im$ is the $2 \times 2$ identity matrix. Stability of the elastic free
energy Eq.~(\ref{elastic}) requires $\mu>0$ and $\lambda + \mu>0$, therefore
$\epsilon>-1$. 

The linear coupling Eq.~(\ref{coupling}) between $\gm$ and $\Qm$
prefers same ``orientation'' (alignment of their eigenvectors) for
these two tensors. Therefore in the ground state, where $\gm$ and $\Qm$
minimize the total free energy, Eq.~(\ref{model}), these two tensors
commute. Using the coordinate system in which $\gm$ and $\Qm$ are
simultaneously diagonal, they can be parameterized as:
\begin{equation}
\gm = \left( \begin{array}{cc}g_1 & 0\\
0 & g_2 \end{array}\right), \hspace{3mm}
\Qm = \left( \begin{array}{cc}S & 0\\
0 & -S \end{array}\right),
\end{equation}
where $g_1$ and $g_2$ are two non-negative eigenvalues of the metric
tensor $\gm$, and $S$ a non-negative magnitude of the nematic order.
Substituting these into free energy $f$, Eq.~(\ref{model}), we find
that it can be written as
\begin{equation}
f = \frac{1}{2}f_g[g_1, g_2,S] + \frac{1}{2}f_S[S],\label{free_energy}
\end{equation}
where
\begin{subequations}
\begin{eqnarray}
f_g[g_1,g_2,S] &=& (2+\,\epsilon)\left((g_1 - \alpha_1)^2 + 
(g_2 - \alpha_2)^2\right)\nonumber\\
&+& 2\, \epsilon\,(g_1 - \alpha_1)(g_2 - \alpha_2), \label{f_g}\\
&=&(1+\epsilon) (g_1+g_2-\alpha_1-\alpha_2)^2,\nonumber\\
&+& (g_1-g_2 - \alpha_1 + \alpha_2)^2, \label{ellipse}\\
 f_S[S] &=& (S^2+\beta)^2,\label{f_S}
\end{eqnarray}
\end{subequations}
and
\begin{subequations}
\begin{eqnarray}
\alpha_1 &=& (S-r),\label{alpha_1}\\
\alpha_2 &=& -(S+r),\label{alpha_2}
\end{eqnarray}
\end{subequations}
with $\alpha_1\ge\alpha_2$, and $f_g$, $f_S$ non-negative definite.

To obtain a mean-field phase diagram, we need to minimize $f$,
Eq.~\ref{free_energy} over nonnegative $g_1$, $g_2$ and $S$ for given
$r$ and $\beta$. Before doing that, however, it is useful to classify
all possible phases according to ground state values of $g_1$, $g_2$
and $S$. It is easy to see that for vanishing nematic order, the free
energy density, Eq.~(\ref{free_energy_0}) reduces to that of an
isotropic polymerized membrane, i.e. Eq.~(\ref{elastic}). An isotropic
polymerized membrane is either in the isotropic-crumpled phase where
$g_1 = g_2 = 0$, or in the isotropic-flat phase, where $g_1 = g_2 >
0$. In other words, a state with $g_1 \neq g_2$ and $S = 0$ cannot be
a stable ground state.  Conversely, anisotropy in the metric tensor,
$g_1\neq g_2$ acts as an external field on the nematic order parameter
$\Qm$, inevitably inducing finite nematic order. Therefore $S$ must be
finite whenever $g_1 \neq g_2$.  Similarly, a nonzero nematic order
acts as an external shear stress tensor, which always induces a finite
shear strain in a flat membrane (Hook's law), leading to elastic
anisotropy with $g_1\neq g_2$. Therefore $g_1$ and $g_2$ cannot be
identical in the presence of a finite nematic order, $S$.  That is, if
$S \neq 0$, the membrane can either be in a nematic-crumpled phase
($g_1 = g_2 = 0$), or a nematic-tubule phase ($g_1 > g_2 = 0$), or a
nematic-flat phase ($g_1 > g_2 \neq 0$).  We list the resulting five
membrane phases in Table.~\ref{phases}.

\begin{table}[!htbp]
\begin{tabular}{|c|c|c|}
\hline\hline $g_1$, $g_2$ and $S$ & $S=0$ 
& $S \neq 0$ \\ \hline $g_1 =g_2=0$ & Isotropic
Crumpled & Nematic Crumpled\\\hline $g_1 = g_2>0$
& Isotropic Flat & Unstable \\\hline $g_1 >g_2 >0$ &
Unstable & Nematic Flat\\\hline $g_1>g_2 = 0$ 
&Unstable& Nematic Tubule\\\hline\hline
\end{tabular}
\caption{Possible phases of a nematic elastomer membrane.} 
\label{phases}
\end{table}

We now minimize the rescaled free energy, Eq.~(\ref{free_energy}) over
$g_1$, $g_2$ and $S$. It is easy to see that if $\beta>0$, $f_S[S]$ in
Eq.~(\ref{f_S}) is minimized by
\begin{equation}
S = 0.
\end{equation}
Substituting this into Eq.~(\ref{f_g}), we find that
$f_g[a_1,a_2,S=0]$ is minimized by
\begin{equation}
g_1 = g_2 =
\left\{\begin{array}{cc}0,& \mbox{for $r \geq 0$}\\ 
                 \sqrt{|r|},&\mbox{for $r<0$}.
\end{array} \right.
\end{equation}
which, as discussed above corresponds to the isotropic-crumpled and
the isotropic-flat phase, respectively.  This situation is illustrated
 in the $\beta-r$ plane in Fig.~\ref{phasediagram}.

For $\beta < 0$, our strategy will be to determine the region of
(meta-)stability for various phases. We first minimize
$f_g[g_1,g_2,S]$, defined by Eq.~(\ref{f_g}), for given $\alpha_1$ and
$\alpha_2$, i.e. for given $S$ and $r$. It is clear that for fixed
values of $\alpha_1$ and $\alpha_2$, a contour $f_g = \mbox{Const.}$
is an ellipse in the $(g_1,g_2)$ plane centered at $(\alpha_1,
\alpha_2)$. Its principle axes are along two diagonal $(1,\pm 1)$
directions, as shown in Fig.~\ref{minimization}. Depending on the
values of $\alpha_1$ and $\alpha_2$, i.e. the center of the ellipse,
the minimizer of $f_g[g_1,g_2,S]$ falls into three categories:
\begin{figure}
\begin{center}
\includegraphics[width=6cm]{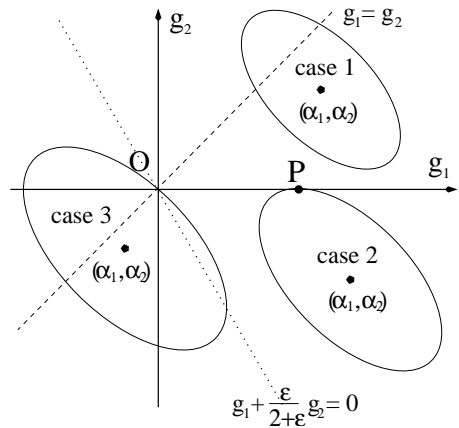}
\caption{Contour plot of $f_g = \mbox{const.}$, an ellipse centered 
  at $(\alpha_1,\alpha_2)$ in the $(g_1,g_2)$ plane, where
  $\alpha_1\leq \alpha_2$, with principle axes along diagonal $(1,1)$
  direction. As discussed in the text, in case 1, $f_g$ is minimized
  by the center of the ellipse, that corresponds to a nematic-flat
  phase. In case 2, it is minimized by the point $T$, where the
  ellipse is tangent with the positive $g_1$ axis. This corresponds to
  a nematic-tubule phase. In case 3, $f_g$ is minimized by the origin
  $g_1 = g_2 =0$, which corresponds to a (nematic or isotropic)
  crumpled phase.}
\label{minimization}
\end{center}
\end{figure}

\begin{enumerate}
\item \begin{math}
\alpha_1 > \alpha_2 > 0: \end{math}

The ellipses are centered within the first quadrant.  It is easy to
see that $f_g$ is minimized by
\begin{equation}
(g_1^*,g_2^*) = (\alpha_1, \alpha_2) = (S-r, -S-r),
\end{equation}
with a minimal value zero. This corresponds to a nematic-flat phase.
Substituting these $g_1^*$ and $g_2^*$ into Eq.~(\ref{free_energy}),
we obtain:
\begin{equation}
f = \frac{1}{2}(S^2+\beta)^2,
\end{equation}
which, when minimized over $S$, gives the magnitude of nematic order
parameter in the nematic-flat phase:
\begin{equation} 
S = \sqrt{|\beta|}.
\end{equation}
The condition of non-negativity of $g_{1,2}$ and $g_2^* = \alpha_2$,
gives the condition of stability of the nematic-flat phase
\begin{eqnarray}
g_2^*&=& -(S+r)\,\\
&=& - ( r + \sqrt{|\beta|})>0,
\end{eqnarray}
which leads to the phase boundary for the nematic-flat phase
illustrated in Fig.~\ref{phasediagram}.

\item $ \alpha_1 + \frac{\epsilon}{2 + \epsilon}\alpha_2>0$ and $\alpha_2 <0$:
  
  In this case, the centers of elliptic free energy contours are
  located outside the first quadrant, and therefore do not correspond
  to physical solution of the metric tensor, characterized by
  $g_{1,2}\ge 0$.  Independent of $\alpha_1$,however, there is
  always one elliptic contour, which is tangent to the positive 
  $g_1$ axis as illustrated in Fig.~(\ref{minimization}). It is
  straightforward to see that this tangent point, $(g_1^*>0, g_2^*
  =0)$ is the minimum of $f_g$ in the physical space of metrics, which
  corresponds to the nematic-tubule phase. At this tangent point, 
  the partial derivative of $f_g$ with respect to $g_1$ vanishes, i.e.
\begin{equation}
\frac{\partial f_g}{\partial g_1}|_{(g_1^*,0)}=
2 (2+ \epsilon )
(g_1^* - \alpha_1 - \frac{\epsilon\,\alpha_2}{2+\,\epsilon} ) = 0,
\end{equation}
which leads to
\begin{eqnarray}
g_1^* &=& \alpha_1 + \frac{\epsilon\,\alpha_2}{2 + \epsilon} \nonumber\\
&=&\frac{2\left( S - \left( 1 + \epsilon  \right) r \right) }{2 + \epsilon }.
\end{eqnarray}
Substituting this $g_1^*$ and $g_2^* =0$ into Eq.~(\ref{free_energy}),
the rescaled total elastic free energy reduces to
\begin{equation}
f = \frac{2\,(1+\epsilon)}{(2+\epsilon)} (S+r)^2 + \frac{1}{2}(S^2 + \beta)^2.
\label{f_NT}
\end{equation}
The equilibrium value of $S$ is determined by minimizing this $f$ with
constraints
\begin{subequations}
\label{constraints}
\begin{eqnarray}
\alpha_2 < 0 &\rightarrow& S > -r,\\
 \alpha_1 + \frac{\epsilon\,\alpha_2}{2 + \epsilon}>0
&\rightarrow& S > (1+\epsilon) r.
\end{eqnarray}
\end{subequations}
For a negative $r$, it is easy to see that as long as
$r<\sqrt{|\beta|}$, there is always a positive value of $S$ that both
minimizes Eq.~(\ref{f_NT}) and satisfies constraints,
Eqs.~(\ref{constraints}).  The case of positive $r$ is more subtle. By
a technically involved, but conceptually straightforward calculation,
we find that for $\beta < -2 - \frac{1}{2 + \epsilon }$ (i.e. to the
left of point $C_1$ in Fig.~\ref{phasediagram}), a nematic tubule
state is stable as long as $(1+\epsilon)r>\sqrt{|2+\beta|}$.  For
$-2(1+\epsilon)/(2+\epsilon)> \beta > -2 - \frac{1}{2 + \epsilon }$
(i.e. to the right of $C_1$, but to the left of $C_2$ in
Fig.~\ref{phasediagram}), we find that a metastable nematic-tubule
state exists as long as
\begin{equation}
\beta + \frac{2(1+\epsilon)}{2+\epsilon}
< -3\left( \frac{(1+\epsilon)r}{2+\epsilon} \right)^{\frac{2}{3}}.
\end{equation}
And finally for $-2(1+\epsilon)/(2+\epsilon)< \beta <0$ (i.e. to the
right of $C_2$ in Fig.~\ref{phasediagram}), nematic tubule phase is
stable if and only if $r<0$.

\item
$\alpha_1+\frac{\epsilon}{2 + \epsilon}\alpha_2<0$ and $\alpha_2<0$: 

In this case, there is no ellipse tangent with the positive $g_1$
axis, i.e. the nematic-tubule phase is not stable. There is however
always one ellipse passing through the origin, $g_1^* = g_2^* =0$ (see
Fig.~\ref{minimization}) that minimizes $f_g$. Depending on the value
of $S$, the corresponding phase may be nematic-crumpled or
isotropic-crumpled. Setting $g_1 = g_2 = 0$ in
Eq.~(\ref{free_energy}), we obtain
\begin{equation}
f = \frac{1}{2}S^4 + (2 + \beta) S^2 + \frac{1}{2}\beta^2 + 2(1+\epsilon) r^2,
\end{equation}
where we have used Eq.~(\ref{alpha_1}) and Eq.~(\ref{alpha_2}).
Minimizing it over $S$, we find the equilibrium value of nematic order
parameter
\begin{equation}
S = \left\{ \begin{array}{cc} 0,& \mbox{if $\beta+ 2 >0$}\\
    \sqrt{|\beta+2|}&\mbox{if $\beta+ 2 <0$}
    \end{array}\right.
\end{equation}
Considering the self-consistent condition that defines case 3, as well
as the definition of $\alpha_1$ and $\alpha_2$, we find that for
negative $\beta$, the isotropic-crumpled phase is stable if and only
if
\begin{equation}
\beta + 2 >0 \hspace{5mm} \mbox{and} \hspace{5mm} r>0.
\end{equation}
On the other hand the nematic-crumpled phase is stable if and only if
\begin{equation}
\beta + 2 <0 \hspace{1mm} \mbox{and} \hspace{1mm}
0<\sqrt{|\beta + 2|} < (1+\epsilon)r.
\end{equation}
\end{enumerate}

We have thereby determined the region of meta-stability for all five
phases admitted by the nematic-elastomer membrane. These results are
summarized in Fig.~\ref{phasediagram}. In the $(\beta,r)$ plane the
small triangular region between three critical points $C_1$, $C_2$ and
$C_3$ is particularly interesting. This region is divided into two
parts by the vertical line $\beta = -2$ passing through $C_3$. In the
left part, both nematic-tubule and nematic-crumpled phases are
metastable. In the right part, both nematic-tubule and
isotropic-crumpled phase are metastable. Consequently, there exists a
first-order transition line between $C_1$ and $C_2$. This line is
illustrated schematically by a dashed curve in
Fig.~\ref{phasediagram}. It connects a second-order
nematic-tubule---to---nematic-crumpled transition boundary on the left
to a second-order nematic-tubule---to---isotropic-crumpled phase
boundary on the right. The exact position of this first-order line,
which can in principle be determined from comparing free energy of
different phases, is not essential to our discussion. Outside the
triangular region, all phase transitions are continuous in our
mean-field analysis.

For completeness, in Table.~\ref{C123} we list the coordinates of the
three critical points, $C_1$, $C_2$ and $C_3$ in the $(\beta,r)$
plane.  It is interesting to note that distances between any two of
these points, and therefore the triangular area, vanishes as
$\epsilon$ becomes large, corresponding to an incompressible limit of
vanishing shear-to-bulk moduli ratio. Due to the same reason, in the
limit of small shear modulus, the discontinuity of first-order
transition line (jump of order parameter or other physical quantities)
vanishes.

\begin{table}[!htbp]
\begin{tabular}{|m{5mm}|m{15mm}|m{15mm}|m{1cm}|}
\hline\hline   & $C_1$  &$C_2$ & $C_3$\\ \hline
$\beta$ & $-2-\frac{1}{2+\epsilon}$ & $-2 + \frac{2}{2+\epsilon}$ 
& $-2$ \\\hline $r$ & $\frac{1}{(1+\epsilon)\sqrt{2+\epsilon}}$&$0$&$0$
\\\hline\hline
\end{tabular}
\caption{Coordinates of three critical points $C_1$, $C_2$ and $C_3$.}
\label{C123}
\end{table}

\section{Phantom Nematic-Tubule Phase}
\label{sec:tubule}
Among all phases that we have identified in the preceeding section,
the nematic tubule phase is the most interesting one. As we have just
found, it is characterized by a nonzero nematic order and a metric
tensor with one positive eigenvalue, i.e.
\begin{equation}
\Qm_0 = \left( \begin{array}{cc}S & 0\\
 0 & -S \end{array}\right), \hspace{5mm}
\mm{g}_0 = \left( \begin{array}{cc} g_1&0
\\0&0\end{array}\right).
\label{gQ_0}
\end{equation}
Even though the ground state configuration of this nematic tubule
phase is identical to that of the tubule phase studied in
Ref.~[\onlinecite{RTtubules}], as we will show in this
section, its elasticity and thermal fluctuations differ qualitatively,
due to the fact that rotational symmetry is {\em spontaneously}
broken, both in the reference space and in the target space.

\subsection{Harmonic Fluctuations}
We first study harmonic thermal fluctuations in the nematic tubule
phase. Without loss of generality we pick the intrinsic (reference
space) membrane axes such that in the ground state the $x$-axis
coincide with the extended direction of tubule, with tubule therefore
crumpled along the $y$-axis. Similarly, our choice of the target space
coordinate system orients the tubule along $\hat{X}$-axis of the
embedding space.  With this choice, the tubule ground-state
conformation is given by
\begin{equation}
\Rv_0(\rv) = \zeta x \hat{X},\label{Rv_0}
\end{equation}
where $g_1 = \zeta^2$ can be calculated by minimizing the total
free energy Eq.~(\ref{model}).

To study thermal fluctuations about the nematic-tubule ground state,
we follow the parameterization introduced in
Ref.~[\onlinecite{RTtubules}] for the study of an intrinsically-anisotropic
tubule. A deviation away from the ground state can be
parameterized by a one-dimensional (scalar) phonon field $u(\rv)$ and
a two-dimension undulation field $\hv(\xv)$, defined by
\begin{equation}
\Rv(\rv) = ( \zeta x + u(\rv))\hat{X} + \hv(\rv). \label{def_uh}
\end{equation}
The corresponding metric tensor, Eq.~(\ref{def_g}), is then given by
\begin{subequations}
\begin{eqnarray}
g^0_{xx} &=& ( \zeta + \px u)^2 + (\px \hv)^2,\\
g^0_{xy} &=& (\zeta + \px u) (\py u) + (\px \hv) \cdot (\py \hv),\\
g^0_{yy} &=& (\py u)^2 + (\py \hv)^2.
\end{eqnarray}
\end{subequations}
The Lagrange strain tensor is defined as half the deviation of 
the metric tensor from its ground state value $\gm_0$, i.e.
\begin{equation}
u_{ab} = \frac{1}{2} ( g_{ab} - g^0_{ab})\label{def_u}.
\end{equation}
Its components are given by
\begin{subequations}
\label{strain_components}
\begin{eqnarray}
u_{xx} &=& \zeta\,(\px u) + \frac{1}{2} (\px u)^2 
+ \frac{1}{2} (\px \hv)^2,\\
u_{xy} &=& \frac{1}{2}(\zeta+\px u)(\py u)
+ \frac{1}{2}(\px \hv) \cdot (\py \hv),\\   
u_{yy} &=& \frac{1}{2}  (\py u)^2 
 + \frac{1}{2} (\py \hv)^2.
\end{eqnarray}
\end{subequations}

Fluctuations of the nematic order parameter $\Qm$ about its 
ground-state value $\Qm_0$ can be parameterized by two variables 
$\tau$ and $\sigma$ defined by:
\begin{equation}
\Qm = \Qm_0 + \delta \Qm
 = \left( \begin{array}{cc}S +  \tau & \sigma \\
\sigma & -S - \tau \end{array}\right). \label{Q_fluct}
\end{equation}
They correspond to longitudinal and transverse excitations of the
nematic order, respectively. Their physical significance can be readily
seen by comparing Eq.~(\ref{Q_fluct}) with the more usual parameterization
of nematic fluctuation in terms of rotation angle $\theta$ and magnitude
fluctuation $\delta S$:
\begin{equation}
\Qm = (S+\delta S)\left( \begin{array}{cc} \cos 2\theta 
& \sin 2 \theta\\\sin 2 \theta & -\cos 2\theta  \end{array}\right). 
\end{equation}
We find that up to linear order of $\delta S$ and $\theta$,
\begin{equation}
\sigma \approx 2 S \theta, \hspace{5mm} \tau \approx \delta S.
\end{equation}
Therefore the longitudinal component $\tau$ describes fluctuations in
the magnitude of nematic anisotropy about $S$, while the transverse
component $\sigma$ is proportional to the rotation of the nematic 
director. 

When expressed in terms of these scalar fields and the Lagrange strain
tensor, the nematic-tubule free energy, Eq.~(\ref{model}), becomes
\begin{eqnarray}
f[\gm, \Qm] &=& \frac{\left( 1 + \epsilon  \right)}{2} \,
     {\left( 2\,r + {\zeta }^2 + 2\,{u_{xx}} + 
         2\,{u_{yy}} \right) }^2,\nonumber\\
&+& \frac{1}{2} {\left( \beta  + 2\,{\sigma }^2 + 
        2\,{\left( S + \tau  \right) }^2 \right) }^2
                + 2\,{\left( \sigma  - 2\,{u_{xy}} \right)}^2
                \nonumber \\
&+& 2\,\left( S - {\zeta }^2/2  
+ \tau  - {u_{xx}} + {u_{yy}} \right)^2. \label{f_tubule}
\end{eqnarray}

The condition that $\gm_0$ and $\Qm_0$, as given by Eq.~(\ref{gQ_0}), 
minimize the elastic free energy density $f[\gm,\Qm]$ guarantees that
in Eq.~(\ref{f_tubule}), terms that are linear in $\tau$ or $u_{xx}$
strictly vanish, and thereby leads to two equations that determine the
ground-state order parameters $S$ and $\zeta$:
\begin{subequations}
\label{eq_condition}
\begin{eqnarray}
-2\,S + 2\,r\,\left( 1 + \epsilon  \right)  + 
  \left( 2 + \epsilon  \right) \,{\zeta }^2 &=&0,\\
4\,S^3 + 2\,S\,\left( 1 + \beta  \right)  - 
  {\zeta }^2&=&0.
\end{eqnarray}
\end{subequations}

Finally, eliminating the strain components $u_{ab}$ in favor of the phonon
($u$) and height undulation ($\hv$) fields using 
Eqs.~(\ref{strain_components}), and keeping only upto quadratic terms,
we obtain the harmonic free energy density of the nematic-tubule:
\begin{eqnarray}
f[\sigma, \tau,u,\hv]
 &=& f[\gm, \Qm]  - f[\gm_0, \Qm_0]\nonumber\\
 &=&    K_{\sigma} (\nabla \sigma)^2    
 +   C_{\sigma} (\sigma - \alpha_{\sigma} \py u)^2 \nonumber\\
&+&     K_{\tau} (\nabla \tau)^2
  +   C_{\tau} (\tau - \alpha_{\tau} \px u)^2 \nonumber\\
&+&     B_u (\px u)^2 + K_u (\py^2 u)^2 \nonumber\\
&+&  B_h (\py \hv)^2 + K_h (\px^2 \hv)^2, \label{f2_tubule}
\end{eqnarray}
where 
\begin{subequations}
\begin{eqnarray}
C_{\sigma}   &=& 2\,\left( 1 + 2\,S^2 + \beta  \right),\\
C_{\tau}     &=& 2\,\left( 1 + 6\,S^2 + \beta  \right),\\
 \alpha_{\sigma} &=& \frac{\zeta}{1 + 2\,S^2 + \beta },\\
 \alpha_{\tau}          &=& \frac{\zeta}{1 + 6\,S^2 + \beta },\\
B_u  &=& 2\,\left( 2 + \epsilon  \right) \,{\zeta }^2,\\
B_h  &=& 4\,S - 2\,{\zeta }^2.
\end{eqnarray}
\end{subequations}

There are several salient features of the nematic-tubule free energy
density, Eq.~(\ref{f2_tubule}), that are worth pointing out. Firstly,
terms that are first-derivative of $\hv$ with respect to the ordered 
direction, $x$, namely $(\px \hv)^2$, do {\em not} appear in $f$. Small
$\px \hv$ corresponds to an infinitesimal rotation in the embedding
space of the tubule's extension axis, that we have arbitrarily chosen 
to point along $\hat{X}$ and therefore must not cost any elastic 
free energy. Therefore absence of the $(\px \hv)^2$ term is a direct 
consequence of the underlying rotational invariance of the 
{\em embedding} space, that a tubule, extended along $\hat{X}$, 
spontaneously breaks.\cite{RTtubules} 
Secondly, $f$ exhibits a linear coupling of the longitudinal
fluctuation of the nematic-order parameter, $\tau$
to the longitudinal strain, $\px u$. This corresponds to the linear 
coupling between $\Qm$ and the metric tensor $\gm$ in the original
model Eq.~(\ref{model}) and encodes the fact
that nematic order induces uniaxial strain and vice versa. Lastly and
most importantly, the transverse fluctuation of the nematic order,
$\sigma$ is ``minimally''\cite{comment_minimalcoupling} coupled to 
the transverse strain component $\py u$, with the coefficient of the
linear coupling $\sigma\py u$ precisely such that it forms a complete
square with two other terms $\sigma^2$ and $(\py u)^2$. This ensures 
that a small and nonzero transverse strain $\py u$ (corresponding to a
reorientation of the uniaxial strain axis) can be completely compensated 
by a uniform perturbation $\sigma$, which corresponds to a global rotation 
of the nematic director within the reference space, so that the overall 
free energy change is exactly zero. This
property is a consequence of the underlying rotational invariance of
the membrane's {\em reference} space, spontaneously broken by the nematic
order, and is therefore expected on general symmetry principles. One
consequence of such coupling is that integration over the nematic
director field, $\sigma$, leads to an effective elastic theory for the 
nematic-tubule, in which the shear term $(\py u)^2$ is absent.
The resulting extra ``softness'' is the essential feature that
 qualitatively distinguishes a nematic-tubule studied here from a 
tubule in an explicitly anisotropic polymerized membrane studied 
by Radzihovsky and Toner.\cite{RTtubules} It reflects the fact 
that in a nematic-tubule the rotational symmetry in the reference
space is only {\em spontaneously} broken, and therefore a simultaneous 
rotation of the nematic order and the metric tensor in the reference 
space does not cost energy. For mechanical stability we have therefore 
added to the free energy Eq.~(\ref{f2_tubule}) two curvature terms, 
$K_u$ and $K_h$, that would otherwise be unimportant for smooth
deformations.

To study thermal undulations of the nematic-tubule it is convenient
to integrate out fluctuations of nematic order parameter, $\tau$ and
$\sigma$, in a direct analogy of integrating out the nematic director 
to study properties of the smectic phase. This leads to an effective
elastic Hamiltonian for the nematic-tubule (in harmonic approximation)
\begin{subequations}
\label{H_harmonic}
\begin{eqnarray}
H_{\mbox{eff}}[u,\hv] &=& \frac{1}{2} \, \int d x d y \left[
 B_u (\px u)^2 + K_u (\py^2 u)^2 \right.  \nonumber\\
 &+&\left. B_h (\py \hv)^2 + K_h (\px^2 \hv)^2 \right],\\
&=& \frac{1}{2} \, \int dq_xdq_y \left[
 (B_u q_x^2 + K_u q_y^4)\,|u(\qv)|^2  \right. \nonumber\\
&+&     \left.  (B_h q_y^2 + K_h q_x^4) \,|\hv(\qv)|^2\right],
\end{eqnarray}
\end{subequations}
where in the last equation we have also expressed the elastic 
Hamiltonian in terms of the Fourier components defined by:
\begin{subequations}
\begin{eqnarray}
u(\qv) &=& \int dxdy e^{-i \qv \cdot \rv} u(\rv) ,\\
\hv(\qv) &=& \int dxdy e^{-i \qv \cdot \rv} \hv(\rv).
\end{eqnarray}
\end{subequations}

From $H_{\mbox{eff}}$ we observe that at the harmonic level, fluctuations 
of $\hv$ field and the phonon field $u$ are decoupled from each other. 
Furthermore, the in-plane phonon-field $u$ elasticity is identical to that
of extensively studied two-dimensional smectic liquid crystal with layers
extended along $y$.\cite{membrane_salditt,membrane_nematicflat}
Therefore we can immediately adopt known results for a two-dimensional 
smectic to fluctuations of the nematic-tubule 
phase.\cite{membrane_salditt,membrane_nematicflat}
For example, for a membrane with intrinsic size $L_x \times L_y$, the bulk 
contribution to mean-squared fluctuations (see below for contribution from 
other modes) for $u$ is given by:
\begin{eqnarray}
\langle u^2 \rangle &=& 
\frac{1}{ B_u} \sqrt{\frac{L_x} { 2 \pi a_u}} 
\psi_u( 2 \pi a_u L_x /L_y^2 )\label{uuhh} \\
& = & \begin{cases}
\frac{1}{\sqrt{2} \pi B_u} \sqrt{\frac{L_x}
{2 \pi a_u}}, & \text{if $L_y^2 \gg 2 \pi a_u L_x$; } \\
\frac{1} {(2 \pi)^2} \frac{1} {B_u } \frac{L_y}{a_u}, 
&\text{if $L_y^2 \ll 2 \pi a_u L_x$ ,} 
\end{cases}
\nonumber
\end{eqnarray}
Therefore, thermal fluctuations of the phonon field, $u$ grow without
bound with membrane's intrinsic size ($L_x$, $L_y$).  This growth of
thermal fluctuations of the $u$ field is a signature of the
qualitative importance of the hitherto neglected nonlinear elastic
terms. 

From Eq.~(\ref{f2_tubule}) we observe that at long length scales
(beyond $\sqrt{K_\sigma/C_\sigma}$) the nematic orientation field,
$\sigma$ is locked by the ``minimal coupling'' term to transverse
strain fluctuations, $\py u$. This, together with dimensional analysis
allows us to relate the asymptotic scaling of the nematic director
field, $\sigma$ to that of the phonon field $u$ through
\begin{equation}
\langle \sigma^2 \rangle \propto \frac{1}{L_y^2} \, \langle u^2 \rangle.
\end{equation}
Despite a strong growth of $\langle u^2\rangle$ appearing in
Eq.~(\ref{uuhh}), we therefore find that the in-plane nematic
orientational fluctuations remains finite in the thermodynamic limit.

From $H_{\mbox{eff}}$ we observe that the undulation field, $\hv$, is
also governed by the same two-dimensional smectic anisotropic
elasticity, except that the roles of $x$ and $y$ axes are
interchanged. At the harmonic level, it is identical to that of a
permanently anisotropic membrane studied in
Ref.~\onlinecite{RTtubules}. As discussed in great detail there, two
related quantities can be defined to describe the tubule geometry at a
finite temperature, tubule radius of gyration, $R_G$ (thickness) and
tubule roughness, $h_{rms}$ as follows:
\begin{subequations}
\label{Rg_def}
\begin{eqnarray}
R_G^2   &=& \langle |\hv(x,0)-\hv(x,L_y)|^2 \rangle,\\
h_{rms}^2 &=& \langle |\hv(0,y)-\hv(L_x,y)|^2 \rangle.
\end{eqnarray}
\end{subequations}
As indicated in Fig.\ref{nematic_tubule} $R_G$ measures the radius of
a typical cross-section of the tubule that is perpendicular to the
nematic order and tubule axis, while $h_{rms}$ measures wandering of
the tubule transverse its backbone.  As discussed by Radzihovsky and
Toner\cite{RTtubules} fluctuations of these two quantities are
strongly affected by the so-called zero modes, that correspond to
Fourier modes of the undulation field $\hv$ with one vanishing wave
vector component.  Adapting these results to our case, we find:
\begin{subequations}
\label{Rg_h}
\begin{eqnarray}
R_G(L_x,L_y) &=& L_y^{1/4} S_R(L_x/\sqrt{L_y}),\\
h_{rms}(L_x,L_y) &=& L_x^{1/2} S_h(L_x/\sqrt{L_y}),
\end{eqnarray}
\end{subequations}
where two crossover functions $S_R$ and $S_h$ satisfy
\begin{eqnarray}
S_R(x) &\rightarrow&    \begin{cases}\frac{1}{\sqrt{x}}& x\rightarrow 0,\\
\mbox{constant}&x \rightarrow \infty; \end{cases}\\
S_h(x) &\rightarrow&    \begin{cases}\mbox{constant}& x\rightarrow 0,\\
x&x \rightarrow \infty. \end{cases}
\end{eqnarray}
For an approximately square size membrane ($L_x \simeq L_x \simeq L$),
we find that $h_{rms}$ scales linearly with $L$, while $R_G \propto
{L}^{1/4}$. Finally, we note that quantities similar to
Eq.~(\ref{Rg_def}) can also be defined for the fluctuations of the
phonon field $u$. Their asymptotic forms are also given by
Eq.~(\ref{Rg_h}), with the role of $x$ and $y$ axes interchanged.

As was found for explicitly-anisotropic tubules\cite{RTtubules}, we
expect that inclusion of the so-far ignored self-avoidance and elastic
nonlinearity (both from in-plane phonon as well as height undulation)
will swell the nematic-tubule and further stabilize it against thermal
fluctuations. In order to perform such a study, we will need a
formalism of elastic free energy for the nematic tubule that includes
all relevant nonlinearities. We derive this nonlinear elasticity in
the next subsection and leave the study of it to a future research.


\subsection{Nonlinear theory for nematic-tubule}
\label{sec:tubule_nl}
Our goal now is to derive the fully rotationally-invariant (in both
the reference and target spaces) free energy describing nematic-tubule. 
This turns out to be simpler within the purely elastic formulation, 
in which, as discussed in Sec.\ref{sec:model}, the in-plane nematic 
orientational degrees of freedom have been formally integrated out. 
The analysis here, then parallels our work on nonlinear elasticity of 
three-dimensional, bulk nematic elastomers.\cite{Xing-Radz}

In such purely elastic approach nematic-tubule's free energy is a 
functional of the deviation of metric tensor $\gm$ from its 
equilibrium values $\mm{g_0}$. Since the elastic free energy density 
should be invariant under arbitrary rotations in both the reference and 
the target spaces, for homogeneous deformations and in two dimensions it 
must be a function of only two scalar functions of the metric 
tensor\cite{comment_g}:
\begin{subequations}
\begin{eqnarray}
S_1 &=&\Tr\gm,\label{S1}\\
S_2 &=&\Tr\gm^2.\label{S2}
\end{eqnarray}
\end{subequations}

In the nematic-tubule phase, we can expand the effective elastic
free energy density around the ground state corresponding to the metric tensor
$\mm{g_0}$, Eq.~(\ref{gQ_0}):
\begin{eqnarray}
f_{\rm eff}[S_1,S_2]
    &=& f_{\rm eff}[S_1^0,S_2^0] + A_1 \,
    \delta S_1 + A_2 \, \delta S_2 \nonumber\\    
    &+& \frac{1}{2} B_1 \delta S_1^2 
    + B_{12} \delta S_1 \delta S_2 
    + \frac{1}{2} B_2 \delta S_2^2\nonumber\\ 
    &+& O(\delta S_i^3),\label{expansion_f}
\end{eqnarray}
where 
\begin{subequations}
\label{def_S}
\begin{eqnarray}
 \delta S_1 &=& \Tr \gm -\Tr \gm_0 = 2\,\Tr\um,\\
 \delta S_2 &=&\Tr \gm^2 -\Tr \gm_0^2 = 4\,\Tr \mm{g_0}\um + 4\Tr \um^2.
\end{eqnarray}
\end{subequations}
As can be checked a posteriori, terms of order $\delta S_i^3$ and
higher are irrelevant at long-scales in the renormalization-group
sense and can therefore be omitted, as we have done above.  

Using Eq.~(\ref{def_S}), we may further express the elastic
free energy density in terms of the Lagrange strain tensor. To proceed, however, 
we note the following relations between 
$\delta S_1$ and $\delta S_2$:
\begin{subequations}
\begin{eqnarray}
\delta S_1 \delta S_2 &=& 2 \zeta^2\,\delta S_1^2
  + \mbox{irrelevant}\nonumber\\
 &=& 8 \zeta^2 (\Tr\um)^2 + \mbox{irrelevant}, \\
\delta S_2^2 &=& 4 \zeta^4 \,\delta S_1^2
   +  \mbox{irrelevant}\nonumber\\
 &=& 16 \zeta^4 (\Tr \um)^2 +  \mbox{irrelevant}.
\end{eqnarray}
\end{subequations}
Substituting these relations, together with Eqs.~(\ref{def_S}) 
into Eq.~(\ref{expansion_f}), we obtain the following nonlinear
elastic free energy density for the nematic tubule:
\begin{eqnarray}
f[\um] &=& f_{\rm eff}[S_1, S_2] -  f_{\rm eff}[S_1^0, S_2^0]\nonumber\\
    &=& 2 (A_1 + 2 \zeta^2 A_2) \Tr \um  + 4 \, A_2 (- \zeta^2 u_{yy}
   + \Tr \um^2 ) \nonumber\\
  &+&  2(B_1 + 4 \zeta^2 B_{12} + 4 \zeta^4 B_2)(\Tr\um)^2.
\end{eqnarray}
Using the fact that $\gm_0$, Eq.~(\ref{gQ_0}) is a true ground state,
together with the observation, from Eqs.~(\ref{strain_components}), that
$\Tr\um$ and $u_{yy}$ are, respectively, linear and quadratic in the
displacement fields, $u$ and $\hv$, we find that the coefficient of
$\Tr\um$ identically vanishes, namely
\begin{equation}
A_1 + 2 \zeta^2 A_2 = 0. 
\end{equation}
To express the elastic free energy in terms of the phonon field $u$
and the undulation field $\hv$, we furthermore note the following two 
identities:
\begin{subequations}
\begin{eqnarray}
\Tr\um &=&  \zeta \px u 
+ \frac{1}{2} (\nabla u)^2
 + \frac{1}{2} (\nabla \hv)^2,\\
- \zeta^2 u_{yy} + \Tr\um^2 &=& 
(\zeta \px u + \frac{1}{2} (\nabla u)^2 
+ \frac{1}{2} (\nabla \hv)^2)^2 \nonumber\\
&-& \frac{1}{2} 
\left( \zeta \py \hv + (\px u) (\py \hv)
 - (\py u) (\px \hv) \right)^2\nonumber\\
&+& \mbox{irrelevant}
\end{eqnarray}
\end{subequations}
Finally, in order to further simplify the notation, we rescale the 
coordinates in the reference space according to:
\begin{equation}
x \rightarrow \zeta^{-1} x, \label{rescaled}
\end{equation}
thereby obtaining our final expression for the nonlinear elastic
free energy density of a nematic tubule:
\begin{eqnarray}
f[u,\hv] &=& \frac{1}{2} B_u\big[(\px u)
 + \frac{1}{2} (\nabla u)^2
 + \frac{1}{2} (\nabla \hv)^2\big]^2 \nonumber\\
&+& \frac{1}{2} B_h\big[\py \hv (1+\px u)
 - (\py u) (\px \hv)\big]^2\nonumber\\
&+& \frac{1}{2} K_u (\py^2 u)^2 
+ \frac{1}{2} K_h (\px^2 \hv)^2,\label{fnl_tubule}
\end{eqnarray}
with the effective bulk elastic moduli given by:
\begin{subequations}
\begin{eqnarray}
B_u &=&  8 \zeta^4 A_2 + 4 \zeta^4 (B_1 + 4 \zeta^2 B_{12}
 + 4 \zeta^4 B_2),\hspace{0.5cm}\\
B_h &=& - 4 \zeta^4 A_2.
\end{eqnarray}
\end{subequations}

\begin{figure}
\label{geometry}
\begin{center}
\includegraphics[width=6cm,height=6cm]{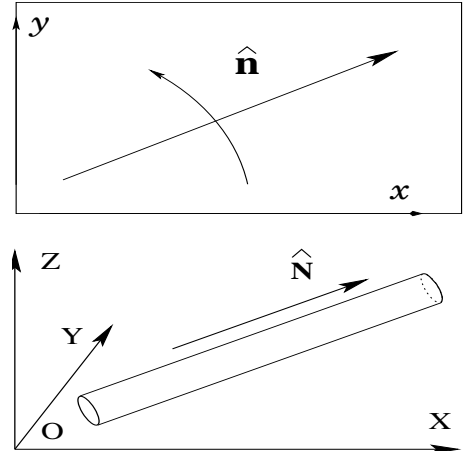}
\caption{A ground state conformation of a nematic tubule can be parameterized
by two unit vectors $\hat{n}_0$ and $\hat{N}_0$, 
Eq.~(\ref{Rv_rotation}), that respectively describe the orientation of
the tubule in the reference and target spaces.
In the reference state Eq.~(\ref{Rv_01}), $\hat{n}_0$ and $\hat{N}_0$ 
are chosen to be $\hat{x}$ and $\hat{X}$ respectively.}
\end{center}
\end{figure}

Despite our ignoring of a number of (at long-scales) subdominant terms, 
it is easy to verify that our final expression for the elastic 
free energy density, Eq.~(\ref{fnl_tubule}), is actually {\em exactly} 
invariant under arbitrary rotations in both the reference and the target 
spaces. To see this we first note that tubule's reference ground-state
conformation, using the rescaled coordinate system Eq.(\ref{rescaled}),
is given by 
\begin{equation}
\Rv_0(\xv) =  \hat{X} x \label{Rv_01}.
\end{equation} 
Under arbitrary rotation both in the reference space and in the 
target space, it becomes
\begin{equation}
\Rv(\xv) = \hat{N}_0 (\hat{n}_0 \cdot \xv) 
\equiv\hat{X} (x + u_0(\xv)) + \hv_0(\xv),
\label{Rv_rotation}
\end{equation}
characterized by target and reference-space unit vectors, $\hat{N}_0$
and $\hat{n}_0$, respectively. This situation is illustrated in 
Fig.~(\ref{geometry}). Without loss of generality, we may parameterize 
these two unit vectors as 
\begin{subequations}
\begin{eqnarray}
\hat{N}_0 &=& \hat{X} \cos \phi + \hat{Y} \sin \phi,\\
\hat{n}_0 &=& \hat{x} \cos \theta + \hat{y} \sin \theta.
\end{eqnarray}
\end{subequations}
From Eq.~(\ref{Rv_rotation}), it is easy to see that 
relative to the reference state $\Rv_0$, Eq.(\ref{Rv_01}), this
pure rotation corresponds to a phonon field $u_0(\xv)$ and an 
undulation field $\hv_0(\xv)$ given by
\begin{subequations}
\begin{eqnarray}
u_0(\xv) &=& \zeta x (\cos \theta \cos \phi -1 ) 
+  \zeta y \sin \theta \cos \phi,\hspace{0.5cm}\\
\hv_0(\xv) &=& \zeta \hat{Y} \sin \phi (x \cos \theta + y \sin \theta).
\end{eqnarray}
\end{subequations}
Substituting these two equations into $f[u,\hv]$, Eq.\ref{fnl_tubule},
we find, that, indeed, as demanded by rotational invariance, the
nonlinear elastic free energy strictly vanishes for an
arbitrary rotation in both the reference and target spaces. 

\section{conclusion}
In this paper, we have presented a Landau theory for polymerized membrane with 
spontaneous in-plane nematic order. We have analyzed the mean-field phase
diagram and studied harmonic thermal fluctuations in the nematic-tubule phase. 
We have also derived a nonlinear elastic free energy for the
nematic-tubule, which will be an essential starting point for future analysis 
of elastic nonlinearities and self-avoiding interaction that we have so 
far ignored. Such a study is necessary in order to understand the physics
of any realistic membrane exhibiting spontaneous nematic tubule phase. 


\begin{acknowledgments}
We thank T. C. Lubensky for discussion and acknowledge support by the NSF MRSEC
DMR-0213918 (LR), DMR-0321848 (LR), David and Lucile Packard Foundation (LR),
as well as NSF DMR02-05858 (XX)and DOE DEFG02-91ER45439(XX).
We also thank Harvard Department of Physics and KITP at UC Santa Barbara,
where part of this work was done, for hospitality.
\end{acknowledgments}




\begin{thebibliography}{10}

\bibitem{polymerized_comment} 
                Also referred as fixed connectivity,
                tethered, or crystal membrane in various references.

\bibitem{membrane_JWS}
                For a review, see {\em Statistical Mechanics of Membranes and Interfaces},
                edited by D.~R.~Nelson, T.~Piran, and S.~Weinberg
                (World Scientific, Singapore, 1989).

\bibitem{NelsonPeliti} 
                D.~R.~Nelson and L.~Peliti, {\em J.~Phys.~(Paris)}
                {\bf 48}, 1085(1987).

\bibitem{membrane_AL}
                J.~A.~Aronovitz and T.~C.~Lubensky, \prl {\bf 60}, 
                2634(1988);
                J.~A.~Aronovitz, L.~Golubovic, and T.~C.~Lubensky,
                {\em J. Phys.(Paris)} {\bf 50}, 609(1989).

\bibitem{membrane_LR}
                P.~LeDoussal and L.~Radzihovsky, \prl
                {\bf 69}, 1209(1992).

\bibitem{membrane_KN} 
                Y.~Kantor and D.~R.~Nelson, \prl {\bf 58}, 2774(1987).

\bibitem{membrane_PKN}
                M.~Paczuski, M.~Kardar, and D.~R.~Nelson,
                \prl {\bf 60}, 2638(1988).

\bibitem{membrane_GD}
                E.~Guitter, F.~David, S.~Leibler, and L.~Peliti,
                \prl {\bf 61}, 2949(1988).

\bibitem{JWS_radzihovsky}
                L. Radzihovsky, to appear in the 2nd Edition of
                {\it the Statistical Mechanics of Membranes and Surface},
                edited by D.~R.~Nelson, T.~Piran, and S.~Weinberg,
                World Scientific, Singapore.

\bibitem{RTtubules} 
                L.~Radzihovsky and J.~Toner, \prl {\bf 75}, 4752(1995); 
                \pre {\bf 57}, 1832 (1998).

\bibitem{membrane_BFT}
                M.~Bowick, M.~Falcioni, and G.~Thorleifsson,
                \prl {\bf 79}, 885(1997); cond-mat/9705059.

\bibitem{GolLub89}
                L. Golubovic and T. C. Lubensky, \prl
                {\bf 63}, 1082(1989).

\bibitem{review_exp} M.~Warner and E.~M.~Terentjev, {\em Prog. Polym. Sci.}
   {\bf 21}, 853(1996), and the reference therein.

\bibitem{review_theory} E. M. Terenjev, {\em J. Phys. Cond. Mat.}
    {\bf 11}, R239(1999).

\bibitem{stress_strainExps}
  H. Finkelmann, I. Kundler, E.M. Terentjev, and M. Warner,
  {\em J. Phys. II} {\bf 7}, 1059(1997);
  G.C. Verwey, M. Warner, and E.M. Terentjev, J.
  {\em Phys. II (France)} {\bf 6}, 1273-1290(1996);
  M. Warner, {\em J. Mech. Phys. solids} {\bf 47}, 1355(1999).

\bibitem{elast_us} 
                T.~C.~Lubensky, R.~Mukhopadhyay, L.~Radzihovsky, and X.~Xing, 
                \pre {\bf 66}, 011702(2002); cond-mat/0112095.

\bibitem{Xing-Radz} 
                X.~Xing and L.~Radzihovsky, 
                {\it Europhys. Lett.} {\bf 61}, 769 (2003); \prl {\bf 90}, 168301(2003).

\bibitem{StenLub} 
                O.~Stenull and T.~C.~Lubensky, {\it Europhys. Lett.} {\bf 61}, 776 (2003);
                \pre {\bf 69}, 021807(2004).

\bibitem{membrane_nematicflat} 
                X.~Xing, R.~Mukhopadhyay, T.~C.~Lubensky, and L.~Radzihovsky, 
                \pre {\bf 68}, 021108 (2003).

\bibitem{comment_soft}
                Similar phenomena take place in other ``soft'' anisotropic
                systems, such as smectic\cite{RTaerogels} and columnar liquid
                crystal phases\cite{RTaerogels}, vortex lattices in putative 
                magnetic superconductors\cite{RT_MSC}, and bulk 
                nematic elastomers\cite{elast_us,Xing-Radz,StenLub}.

\bibitem{comment_nematicflat}
                Stability of an isotropic-flat phase at finite temperature 
                is ensured by length scale dependent bending modulus that
                is infinitely enhanced by thermal fluctuations. Therefore 
                this finding of a finite, length-scale independent
                bending-rigidity modulus for flat nematic elastomer
                membranes suggests that this nematic-flat phase might
                indeed be unstable to thermal fluctuations. However, we have
                also shown that for $D<3$, new {\em in-plane} elastic 
                nonlinearities which is not included in our analysis near 
                $D=4$, become qualitatively important, and can very well
                lead to the sufficient enhancement of the bending rigidity 
                sufficient to stabilize a nematic-flat phase of a 2D membrane. 
                At the moment, this question remains unresolved.

\bibitem{Podgornik}
                R.~Podgornik, \pre {\bf 52}, 5170(1995). 

\bibitem{comment_defects}
                However, topological defects in these two systems are 
                qualitatively different.


\bibitem{comment_phantom}
                As the word ``phantom'' suggests, we will completely ignore 
                the effects of self-avoiding interaction. It should be made 
                clear that the reason we ignore self-avoidance in this work is 
                not its unimportance. On the contrary, past research experience
                in this field shows that except for the flat phase, self-avoiding 
                interaction is qualitatively important to the long length scale 
                properties of polymerized membranes. The primary object of this paper 
                is to study the global phase diagram of a liquid-crystalline tethered  
                membrane. We believe that self-avoidance does not destroy, but 
                rather modifies, various phases that we will identify in this paper. 
                Therefore we will leave the technically involved study of
                self-avoiding interaction to a separate publication.

\bibitem{RTaerogels} 
                L.~Radzihovsky and J.~Toner, \prb {\bf 60} 206(1999);
                \prl {\bf 78} 4414(1997).

\bibitem{RT_MSC}
                L.~Radzihovsky, A.~M.~Ettouhami, K.~Saunders, and J.~Toner,
                \prl {\bf 87}, 027001 (2001).

\bibitem{membrane_random} 
                L.~Radzihovsky and D.~R.~Nelson, \pra{\bf 44}, 3525(1991),
                {\it Europhys. Lett.} {\bf 16}, 79(1991), 
                \pra, {\bf 46} 7474(1992);
                D.~C.~Morse, T.~C.~Lubensky and G.~S.~Grest, \pra {\bf 45},
                R2151(1992); D.~C.~Morse, T.~C.~Lubensky, \pra {\bf 46},
                1751(1992).

\bibitem{Safinya_Unpublished}
                L.~S.~Hirst and C.~R.~Safinya, unpublished.


\bibitem{comment_Q}
                Such an intrinsic nematic order parameter $\mm{Q}$ may, for
                example, characterize a spontaneous lipid tilt-order or 
                alignment of liquid crystal polymers making up a nematic
                elastomer membrane. It should be intuitively clear that the
                component of such orientational order normal to the membrane, 
                should have no anisotropic effects on the elasticity of the
                tethered membrane. Consequently $\mm{Q}$ the nematic order
                parameter is a well-defined reference-space rank-two tensor
                that can couple to membrane's metric tensor $\mm{g}$.
                This contrasts qualitatively with distinct physical
                realization of, for example, a tethered membrane fluctuating 
                inside a nematic liquid crystal. In this latter case the 
                appropriate nematic order parameter is a tensor in the 
                embedding space that can only couple to tangent
                vectors of a membrane. 


\bibitem{LC_deGennes}
                P~G.~de~Gennes and J.~Prost, {\it The Physics of Liquid Crystals},
                Clarendon Press, Oxford, 1993.
                                


\bibitem{comment_crumpling}
                This crumpling transition is second order at a mean field level
                 and is supported by numerical simulation. RG analysis, however, 
                 shows that it may be discontinuous.

\bibitem{comment_minimalcoupling}
                This is mathematically identical to what's called a ``minimal''
                coupling of a charged-field and the corresponding gauge field in 
                gauge theories that ensures gauge invariance, with the identification
                of $\sigma$ with the gauge field, $u$ the phase of the charged 
                field and $1/\alpha_\sigma$ the charge. 


\bibitem{membrane_salditt}
                 T.~Salditt, I.~Koltover, J.~O.~R\"{a}dler, and C.~R.~Safinya, 
                 \pre {\bf 58}, 889(1998).

\bibitem{comment_g}
                This is because in two dimensions the metric tensor 
                $\gm$ is a  $2 \times 2$ matrix and has only two 
                rotationally-invariant independent degrees of freedom, 
                i.e. two eigenvalues.



\end{thebibliography}



\end{document}